\documentclass[10pt]{iopart}
\usepackage{graphicx}
\usepackage{amssymb}
\usepackage{array}
\usepackage{color}
\usepackage{enumerate}
\usepackage{bm}

\begin{document}

\title[]{An emerging global picture of heavy fermion physics}

\author{Yi-feng Yang$^{1,2,3}$}

\address{$^1$Beijing National Laboratory for Condensed Matter Physics and Institute of Physics, Chinese Academy of Sciences, Beijing 100190, China}
\address{$^2$University of Chinese Academy of Sciences, Beijing 100049, China}
\address{$^3$Songshan Lake Materials Laboratory, Dongguan, Guangdong 523808, China}
\ead{yifeng@iphy.ac.cn}
\vspace{10pt}

\begin{abstract}
Recent progresses using state-of-the-art experimental techniques have motivated a number of new insights on heavy fermion physics. This article gives a brief summary of the author's research along this direction. We discuss five major topics including: (1) Development of phase coherence and two-stage hybridization; (2) Two-fluid behavior and hidden universal scaling; (3) Quantum phase transitions and fractionalized heavy fermion liquid; (4) Quantum critical superconductivity; (5) Material-specific properties. These cover the most essential parts of heavy fermion physics and lead to an emerging global picture beyond conventional theories based on mean-field or local approximations.
\end{abstract}

%
\vspace{2pc}
\ioptwocol

\section{Introduction}
Heavy fermion systems, a subset of rare earth and actinide intermetallics with large quasiparticle effective mass, exhibit many exotic correlated phenomena such as non-Fermi liquid and unconventional superconductivity \cite{Stewart2001RMP,Pfleiderer2009RMP}. Underlying all their unusual properties is the delocalization of $f$ moments with temperature or external tuning parameters through collective hybridization with conduction electrons, whose exact mechanism remains to be clarified \cite{Yang2016RPP,Lonzarich2017RPP}. Most people believe that its microscopic physics is contained in the following Kondo-Heisenberg model:
\begin{equation}
H=\sum_{{\bf k}\sigma}\epsilon_{\bf k}c_{{\bf k}\sigma}^\dagger c_{{\bf k}\sigma} + J_{\rm{K}}\sum_i \mathbf{s}_{i}\cdot\mathbf{S}_{i}+ J_{\rm{H}}\sum_{\langle ij\rangle}\mathbf{S}_{i}\cdot\mathbf{S}_{j},
\end{equation}
where the first term describes free conduction electrons, the second one describes their Kondo coupling ($J_{\rm{K}}>0$) with local $f$ spins ($\mathbf{S}_{i}$), and the last one is a Heisenberg term between nearest-neighbor $f$ spins, which is typically induced by the Kondo coupling due to the Rudeman-Kittel-Kasuya-Yosida (RKKY) mechanism but written here explicitly for clarity. 
\begin{figure}[b]
\centering\includegraphics[width=0.4\textwidth]{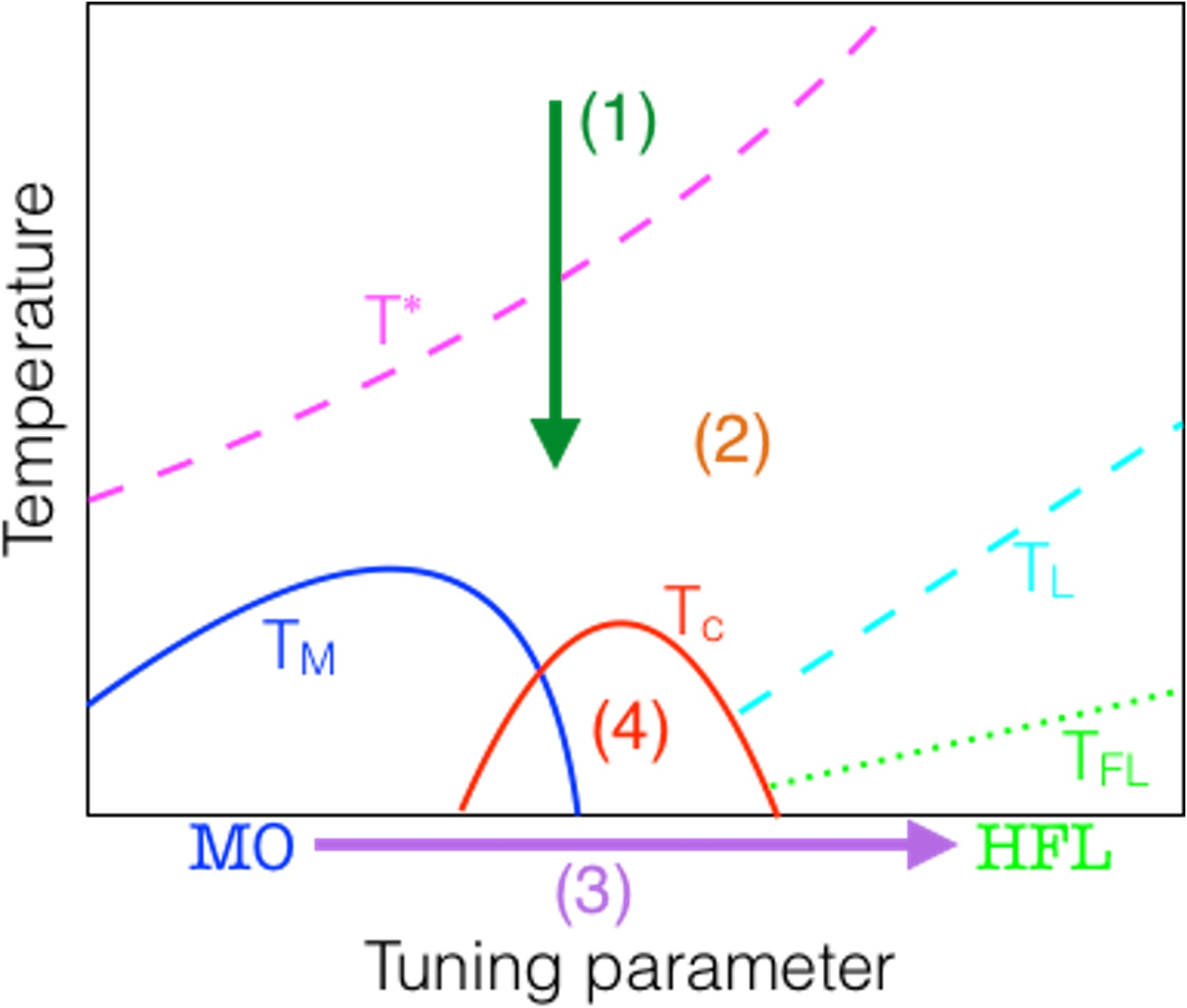}
\caption{A generic phase diagram of the Kondo lattice systems with temperature and external tuning parameters (pressure, magnetic field, doping, etc), showing several regions of different physics: (1) the development of phase coherence at high temperatures; (2) the two-fluid region below the coherence temperature $T^*$; (3) the quantum phase transition from a magnetic order (MO) to a heavy Fermi liquid (HFL); (4) the superconductivity near a quantum critical point (QCP). $T_M$ and $T_c$ mark the magnetic and superconducting transition temperatures, respectively. $T_L$ and $T_{FL}$ mark the crossover temperatures for the full delocalization of $f$ moments (see section \ref{secTFM}) and the Fermi liquid ground state, respectively.}
\label{fig1}
\end{figure}

Conventional theories are mostly based on certain mean-field or local approximations of the above model Hamiltonian \cite{Coleman2015book}, but recent experiments using state-of-the-art techniques have revealed unexpected features that demand more sophisticated explanations involving its dynamical and nonlocal spatial correlations. In this short article, we try to introduce a global picture of heavy fermion physics that emerges from latest experimental and theoretical progresses. It is not a comprehensive review of the whole field but rather a brief summary of the author's study along the above line of thought, so only some most related references are cited. Five major topics will be discussed: 
\begin{enumerate}[(1)]
\item Development of phase coherence.
\item Two-fluid phenomenology.
\item Quantum phase transitions.
\item Quantum critical superconductivity.
\item Material-specific properties.
\end{enumerate}
The first four topics are illustrated in the tentative phase diagram Fig.~\ref{fig1} and their meanings will be discussed in more details in the following sections.

\section{Development of phase coherence}\label{secPC}
The coherence temperature $T^*$ marks one of the most important temperature scales in heavy fermion physics \cite{Yang2016RPP}. It can usually be estimated from the resistivity maximum that separates an insulating-like region due to incoherent Kondo scattering at higher temperatures and a metallic region at lower temperatures \cite{Petrovic2001JPCM}. Many physical quantities exhibit anomalous behaviors around $T^*$ (see section \ref{secTFM}) \cite{Yang2008Nature,Ohish2009PRB,Mo2012PRB}. It is thus important to clarify the microscopic origin of $T^*$. In the mean-field theory, $T^*$ marks the onset of hybridization, which opens both direct and indirect gaps simultaneously \cite{Coleman2015book}. However, this simple picture was questioned recently by the angled-resolved photoemission (ARPES) experiment in CeCoIn$_5$, which reported the bending of conduction bands (an indication of hybridization) already at $T^\dagger\approx120\,$K \cite{Chen2017PRB}, which is well beyond the coherence temperature $T^*\approx 50\,$K estimated from the transport and magnetic measurements \cite{Petrovic2001JPCM}. Later  ultrafast optical pump-probe spectroscopy measurement observed different signatures at $T^\dagger$ and $T^*$ in the relaxation rate ($\gamma$) of excited quasiparticles \cite{Liu2020PRL}. $\gamma$ is almost constant above $T^\dagger$ and starts to decrease at lower temperatures, but below $T^*$, its temperature variation becomes fluence-dependent. It has been argued that this difference reflects the opening of direct and indirect hybridization gaps at $T^\dagger$ and $T^*$, respectively, suggesting a two-stage hybridization process due to phase fluctuations beyond the mean-field theory. Similar behavior was then confirmed in a quite different compound, the ferromagnetic Kondo lattice CeRh$_6$Ge$_4$ \cite{Pei2021PRB}, implying that it is a universal property of heavy fermion systems. It is therefore important to understand how the coherence is developed and eventually  established with lowering  temperature.

The above experimental observations motivated a phase coherence scenario for describing the heavy fermion physics beyond the mean-field picture \cite{Dong2022PRB}. It is formulated based on the Abrikosov pseudofermion representation of local spins, $\mathbf{S}_{i}=\sum_{\eta\gamma}f_{i\eta}^{\dag}\frac{\bm{\sigma}_{\eta\gamma}}{2}f_{i\gamma}$. Using the Hubbard-Stratonovich decomposition, the Kondo-Heisenberg model gives the Lagrangian:
\begin{equation}
\mathcal{L}=\sum_{i}\frac{J_{\rm K}\left\vert V_{i}(\tau)\right\vert ^{2}}{2}+\sum_{\left\langle ij\right\rangle}\frac{J_{\rm H}\left\vert \chi_{ij}(\tau)\right\vert^{2}}{2}+\mathcal{L}_{cf}, \label{eq1}
\end{equation}
where $V_i$ and $\chi_{ij}$ are two auxiliary fields describing onsite hybridizations from the Kondo coupling and intersite magnetic correlations from the Heisenberg interaction due to the RKKY mechanism, respectively. $\mathcal{L}_{cf}$ takes a bilinear form of conduction electrons and pseudofermions which interact through the two fluctuating auxiliary fields. To avoid the sign problem, a static approximation can be employed by taking $V_i(\tau)\rightarrow V_i$ and $\chi_{ij}(\tau)\rightarrow\chi_{ij}$, which ignores temporal fluctuations of the auxiliary fields but takes full account of their spatial fluctuations and statistical distributions \cite{Dong2021PRB}.

\begin{figure}[t]
\centering\includegraphics[width=0.48\textwidth]{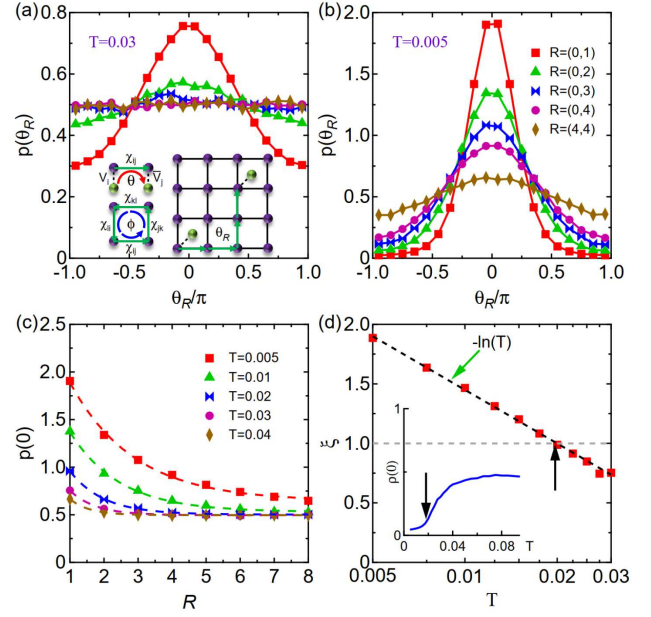}
\caption{Development of phase coherence on the Kondo lattice \cite{Dong2022PRB}. (a),(b) Comparison of the probabilistic distribution of $\theta_R$ for different paths at two temperatures $T=0.03$ and 0.005. The inset of (a) illustrates how the gauge-invariant phases are defined. (c) Exponential decay of $p(\theta_R=0)$ with $R$ for different temperatures. The dashed lines are the exponential fit. (d) The subtracted correlation length $\xi$ as a function of temperature, showing logarithmic development with lowering temperature. The inset plots the conduction electron density of states (DOS) at the Fermi energy. The black arrows mark the temperature where $\xi=1$ and the pseudogap turns into a full gap. The finite DOS inside the gap comes from numerical errors due to artificial broadening. Figure adapted from \cite{Dong2022PRB}. Copyright 2022 American Physical Society.}
\label{fig2}
\end{figure}

The fermionic degrees of freedom $c$ and $f$ can then be integrated out, resulting in the probabilistic distribution of the auxiliary fields only, $p(V_{i},\chi_{ij})=Z^{-1}\rme^{-S_{\rm eff}}$, where $Z$ is the partition function serving as the normalization factor. The effective action $S_{\rm eff}$ is invariant under the gauge transformation $V_{i}\rightarrow V_{i}\rme^{\rmi\beta_{i}}$, $\chi_{ij}\rightarrow\rme^{-\rmi\left(  \beta _{i}-\beta_{j}\right)  }$, so we may define two gauge-invariant phases \cite{Dong2022PRB}:
\begin{eqnarray}
F_{i}  & \equiv\chi_{ij}\chi_{jk}\chi_{kl}\chi_{li}=\left\vert F_{i}%
\right\vert \rme^{\rmi\phi_{i}},\nonumber\\
B_{ij}  & \equiv V_{i}\chi_{ij}\overline{V}_{j}=\left\vert B_{ij}\right\vert
\rme^{\rmi\theta_{ij}},
\end{eqnarray}
where $\phi_{i}$ describes the flux in a plaquette and $\theta_{ij}$ denotes the phase of the hybridization bond $B_{ij}$ between nearest-neighbor sites $ij$ as illustrated in the inset of Fig.~\ref{fig2}a. The latter can be extended to any path on the lattice:
\begin{eqnarray}
\theta_{R}&\equiv\theta_{i_0 i_1}+\theta_{i_1 i_2}+\cdots+\theta_{i_{R-1}i_R}\,\, {\rm mod}\,\, 2\pi  \nonumber\\
&={\rm Im}\ln\left(V_{i_0}\chi_{i_0 i_1}\chi_{i_1 i_2}\cdots \chi_{i_{R-1}i_R}\overline{V}_{i_R}\right),
\end{eqnarray}
where $i_0i_1i_2...i_R$ denotes a path of length $R$ linking two end sites at $\textbf{r}_{i_0}$ and $\textbf{r}_{i_R}\equiv\textbf{r}_{i_0}+\textbf{R}$. $\theta_{R}$ is also gauge-invariant  and reflects the correlation of two hybridization fields mediated by intersite magnetic correlations along the path.

Figures \ref{fig2}a and \ref{fig2}b compare the distribution $p(\theta_R)$ for a number of paths at two different temperatures for a large $J_{\rm K}$ with a Kondo insulating ground state. For the shortest path, both exhibit a maximum at $\theta_R=0$, indicating the presence of short-range correlations at both temperatures. But for $T=0.03$, the distribution turns uniform rapidly as $R$ increases; for $T=0.005$, the maximum is reduced but exists for all $R$. The phase correlation is therefore long-range at $T=0.005$. In Fig.~\ref{fig2}c, we plot $p(\theta_R=0)$ as a function of $R$ for different temperatures. Quite remarkably, all data can be fitted with the exponential function $p(\theta_R=0) =A \rme^{- R/\xi} +B$ (dashed lines), where $\xi$ is a characteristic phase correlation length. As shown in Fig.~\ref{fig2}d, it is less than $1$ (in the unit of lattice parameter) at high temperatures but increases logarithmically (dashed line) with lowering temperature. Around the same temperature, the conduction electron density of states also changes from a pseudogap (direct hybridization gap) to a full gap (indirect hybridization gap). This is exactly the two-stage hybridization, which has also been obtained in exact determinant quantum Monte Carlo (DQMC) calculations on the half-filled periodic Anderson model \cite{Hu2019PRB}.

\begin{figure*}[t]
\centering\includegraphics[width=1.0\textwidth]{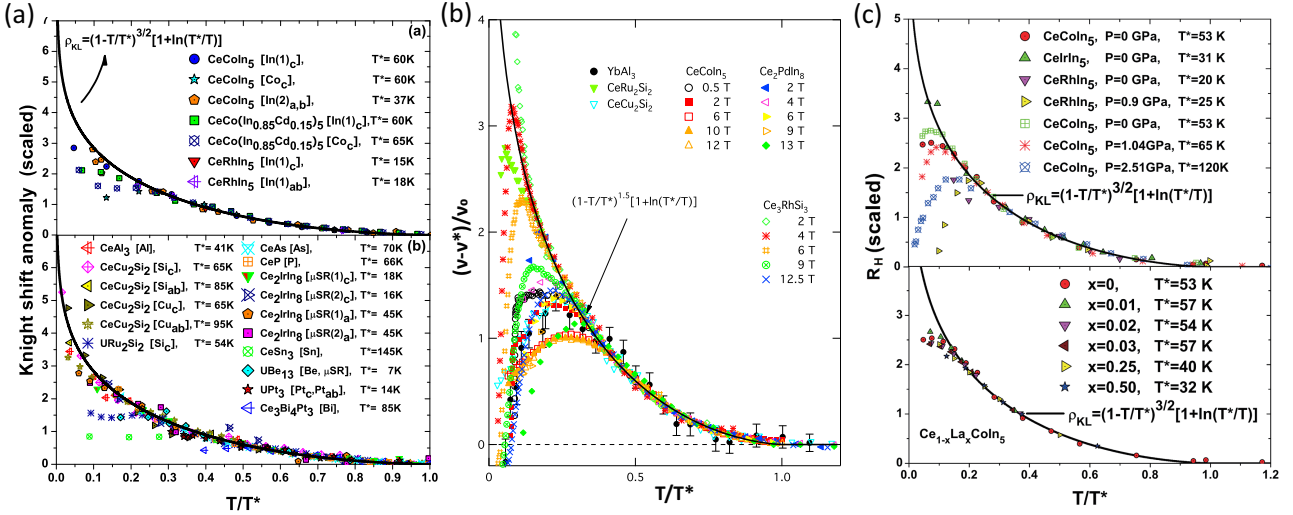}
\caption{Universal temperature scaling predicted by the two-fluid model for (a) the Knight shift anomaly \cite{Yang2008PRL}, (b) the Nernst coefficient \cite{Yang2020PRR}, and (c) the Hall coefficient \cite{Yang2008PRL} in heavy fermion compounds. Figure adapted from \cite{Yang2008PRL,Yang2020PRR}. Copyright 2008 and 2020 American Physical Society.}
\label{fig3}
\end{figure*}

Although it has yet to be extended to more general cases, the above results already provide a novel picture for understanding the heavy electron emergence. One may imagine that the conduction electrons toddle on a phase fluctuating background. Every now and then, they temporarily bind to some local spins and form short-lived composite particles, but the latter can only propagate coherently on the lattice and grow into well-defined heavy quasiparticles when a long-range phase correlation of the hybridization fields is established with the assistance of intersite magnetic correlations. This distinguishes the Kondo lattice physics from the single-impurity Kondo physics, and questions the validity region of the dynamical mean-field approach where spatial correlations are included in some obscure way through its self-consistency procedure \cite{Georges1996RMP,Hu2020PRR}. A two-fluid model might be derived by analogy with that of the superconductivity \cite{Bardeen1958PRL}. In reality, crystal field effect may also play a role at high temperatures and complicate the experimental observations \cite{Jang2020PNAS}. In any case, our predicted two-stage hybridization and the gradual development of the lattice phase coherence should be present in all Kondo lattice systems and may be examined in more future experiments.

\section{Two-fluid phenomenology}\label{secTFM}
Below the coherence temperature $T^*$, the systems exhibit unusual two-fluid behavior \cite{Yang2016RPP,Yang2011JPCM}. This was first observed in La-doped CeCoIn$_5$ for the susceptibility and specific heat \cite{Nakatsuji2004PRL} and later extended to other physical properties in many heavy fermion compounds \cite{Curro2004PRB,Yang2008PRL,Yang2009PRL,Yang2009PRB,Bauer2011PNAS,Yang2012PNAS,Shirer2012PNAS,Yang2013PRB,Yang2015Wuli,Yang2015PRB,Yang2020PRR}. The two fluids are a heavy electron fluid of the screened $f$ moments and a spin liquid fluid formed of the unscreened ones \cite{Yang2012PNAS}. This is a quite intuitive picture considering that the $f$ moments are only partially screened in a broad intermediate temperature range. All physical quantities should in principle contain contributions from both components. For example, the Knight shift and the magnetic susceptibility may be decomposed as
\begin{eqnarray}
K&=&K_0+A\chi_l+B\chi_h\nonumber\\
\chi&=&\chi_l+\chi_h,
\end{eqnarray}
where $\chi_l$ and $\chi_h$ denote the local and heavy electron contributions, respectively. For $T>T^*$, all $f$ moments are presumably localized, so we have $\chi_h=0$ and  $K=K_0+A\chi$. But for $T<T^*$, heavy electrons emerge, the linear relationship between $K$ and $\chi$ is violated, and we have a Knight shift anomaly,
\begin{equation}
K_{\rm a}=K-K_0-A\chi=(B-A)\chi_h,
\end{equation}
which is exactly what nuclear magnetic resonance (NMR) measurements observed \cite{Curro2001PRB,Curro2009RPP}. The two-fluid behavior can even persist inside the superconducting phase \cite{Yang2009PRL}, and thus provides a phenomenological explanation for the microscopic coexistence of local moment AFM order and superconductivity in CeRhIn$_5$ under pressure \cite{Park2006Nature}.

Quite unexpectedly, the heavy electron fluid exhibits universal scaling with temperature \cite{Yang2008PRL}: 
\begin{equation}
\chi_{h}\propto f_0\left(1-\frac{T}{T^*}\right)^{3/2}\left(1+\ln\frac{T^*}{T}\right)
\end{equation}
where $f(T)\equiv f_0(1-T/T^*)^{3/2}$ represents the fraction of hybridized $f$ moments and takes values within $[0,1]$. $f(T)$ may be viewed as an ``order parameter" measuring the $f$ electron itinerancy, and the prefactor $f_0$ controls the effectiveness of the hybridization \cite{Yang2012PNAS}. As shown in Fig.~\ref{fig3}, the predicted universal scaling has been widely observed in the Knight shift anomaly \cite{Yang2008PRL} and the Nernst coefficient \cite{Yang2020PRR} in many heavy fermion compounds, and the Hall coefficient in some Ce-based compounds \cite{Yang2008PRL,Yang2013PRB} over a wide intermediate temperature range. This universality reflects a generic mechanism underlying the complicated heavy fermion physics. Systematic experimental analyses suggest that $T^*$ is proportional to the RKKY interaction \cite{Yang2008Nature}, so the heavy electron emergence should be closely associated with intersite magnetic correlations as discussed in section \ref{secPC} rather than the widely-believed Kondo scale predicted by the mean-field theory \cite{Burdin2009PRB}.

Note that what is not included in the above scaling formula is that it should break down at low temperatures if other physics intervenes or all $f$ moment become delocalized when $f(T)$ reaches one for $f_0>1$. This latter observation motivated a surprisingly simple two-fluid picture for understanding the ground state properties of the Kondo lattice \cite{Yang2012PNAS,Yang2014PNASa,Yang2014PNASb,Yang2017PNAS}. For $f_0<1$, $f(T)$ is always smaller than one at all temperatures, so there must exist residual unscreened $f$ moments that persist at very low temperatures to form either a spin liquid or a magnetic ground state. Above the ordering temperature $T_{M}$, the electron Fermi surface is expected to be of intermediate size due to the partial itinerancy, consistent with ARPES measurements \cite{Chen2017PRB}. Near the transition, magnetic fluctuations may drive heavy electrons back into local moments (called relocalization) \cite{Shirer2012PNAS,Warren2011PRB}, so that the Fermi surface might be reduced deep inside the ordered phase to resolve the different observations between de Haas-van Alphen (dHvA) and ARPES measurements in some compounds like CeRhIn$_5$ \cite{Shishido2002JPSJ,Harrison2004PRL,Chen2018PRL}. For $f_0>1$, all $f$ moments become itinerant below a finite temperature $T_L$ defined by $f(T_L)=1$, and the ground state is a heavy Fermi liquid with a large Fermi surface. $f_0=1$ marks a delocalization transition at zero temperature. If the heavy Fermi liquid is magnetically unstable, the delocalization transition appears inside the ordered phase of itinerant magnetism (type II). If the spin interaction or lattice geometry is highly frustrated, the local moment order could be destroyed before the delocalization transition, then we will have an intermediate paramagnetic region of coexisting spin liquid and heavy electrons (type III). Otherwise, the magnetic and delocalization transitions coincide at the same point (type I), as illustrated in Fig.~\ref{fig1}. Many of the important temperature scales including the superconducting dome can be well fitted within this two-fluid framework \cite{Yang2014PNASa,Yang2014PNASb}. The quantum criticality is determined by the interplay of magnetic and hybridization fluctuations \cite{Yang2017PNAS}.

The two-fluid phenomenology and its predicted universal scaling have been continuously examined by a large amount of different kinds of measurements \cite{Yang2016RPP}. Model calculations using various numerical methods have also been applied to understand the two-fluid ``order parameter", the coherence temperature scale, and the scaling formula \cite{Barzykin2006PRB,Zhu2011PRB,Choi2012PRL,Jiang2014PRB,Xie2015PRB,Jiang2017PRB,Yang2015PP}, but a microscopic derivation of the two-fluid phenomenology has yet to be achieved.

\section{Quantum phase transitions}\label{secQPT}
The above three situations provide a simple classification of AFM quantum phase transitions observed experimentally in CeCu$_2$Si$_2$ (type II), CePdAl (type III), and CeRhIn$_5$ or YbRh$_2$Si$_2$ (type I), respectively \cite{Yang2017PNAS,Yang2020SCPMA}. Type II exhibits the usual spin-density-wave type quantum criticality described by the Hertz-Millis-Moriya theory \cite{Hertz1976PRB,Millis1993PRB,Moriya1995JPSJ,Lohneysen2007RMP}. Tremendous efforts have been made to understand type I with coincident QCP \cite{Si2001Nature,Coleman2001JPCM,Gegenwart2008NatPhys,Stockert2011ARCMP,Abrahams2012PNAS,Steglich2014PM}. Type III is rare but may appear in frustrated Kondo lattice systems such as CePdAl with a quasi-Kagome structure, in which an intermediate non-Fermi liquid phase was recently reported under pressure and magnetic field tuning \cite{Zhao2019NatPhys,Zhang2018PRB}. Both I and III involve the destruction of local moment AFM orders and demand a unified explanation beyond conventional pictures.

A microscopic theory was lately developed based on the Schwinger boson representation of local spins \cite{Wang2021PRB}. Compared to the pseudofermion representation, the boson representation is more suitable to describe magnetic fluctuations in the Kondo-Heisenberg model \cite{Arovas1988PRB,Pepin2005PRL,Rech2006PRL,Komijani2018PRL,Komijani2019PRL,Wang2020PRB}. We get the Lagrangian:
\begin{equation}
\mathcal{L}=\mathcal{L}_{c}+\mathcal{L}_{b}+\mathcal{L}_{\chi}+\frac{1}{\sqrt{N}}\sum_{i\alpha a}\left(b_{i\alpha}^\dagger c_{ia\alpha}\chi_{ia}+\rm{H.c.}\right),
\end{equation}
where $\mathcal{L}_{c}=\sum_{{\bf k}\alpha a}c_{{\bf k}\alpha a}^\dagger (\partial_\tau+\epsilon_{\bf k})c_{{\bf k}\alpha a}$ gives the electron dispersion, $\mathcal{L}_b$ describes the bosonic spinons from the Heisenberg interaction, and $\mathcal{L}_\chi=\sum_{ia}|\chi_{ia}|^2/J_{\rm K}$ describes the fermionic holons from Hubbard-Stratonovich decomposition of the Kondo coupling \cite{Parcollet1997PRL}. 

\begin{figure}[t]
\centering\includegraphics[width=0.5\textwidth]{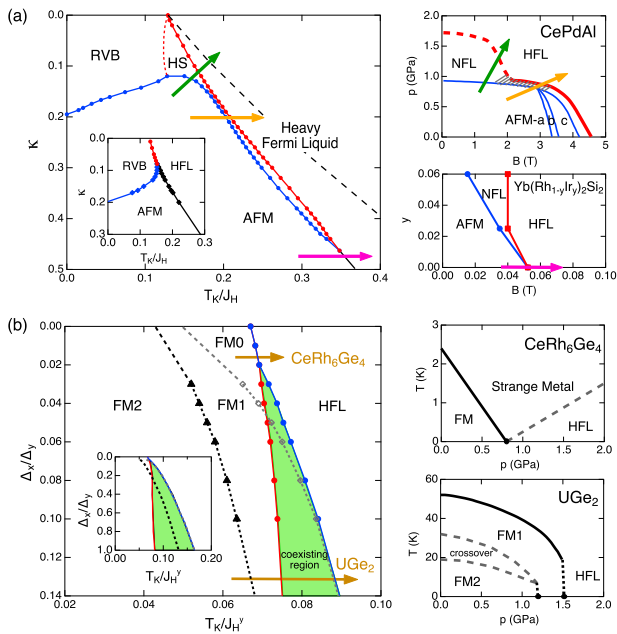}
\caption{Theoretical phase diagrams of the nonlocal Schwinger boson approach for (a) AFM Kondo lattice \cite{Wang2021PRB} compared to experiments CePdAl \cite{Zhao2019NatPhys} and Yb(Rh$_{1-y}$Ir$_y$)$_2$Si$_2$ \cite{Friedemann2009NatPhys}, with $\kappa$ denoting the strength of quantum fluctuations; (b) FM Kondo lattice \cite{Wang2022SCPMAb} compared to CeRh$_6$Ge$_4$ \cite{Shen2020Nature} and UGe$_2$ \cite{Aoki2019JPSJ}, with $\Delta_x/\Delta_y$ denoting the anisotropy of intersite magnetic correlations. The green coexisting region indicates a first-order transition with both FM and HFL solutions. $T_{\rm K}$ and $J_{\rm H}$ ($J_{\rm H}^y$) are the single-ion Kondo temperature and Heisenberg interaction (isotropic for the AFM model and along $y$-axis for the FM model), respectively. Figure adapted from \cite{Wang2021PRB,Wang2022SCPMAb}. Copyright 2021 American Physical Society and 2022 Science China Press.}
\label{fig4}
\end{figure}

The model can be solved under one-loop approximation taking into consideration nonlocal spatial correlations or momentum-dependent self-energies in the large-$N$ approximation \cite{Wang2021PRB}. This gives the global phase diagram on the square lattice in Fig.~\ref{fig4}a. For weak quantum fluctuations (large $\kappa$), we find a direct transition between the AFM and the HFL, showing deconfined quantum criticality as reported probably in YbRh$_2$Si$_2$ \cite{Pfau2012Nature}. For strong quantum fluctuations possibly due to low dimensionality, large spin/orbital degeneracy, or magnetic frustration \cite{Coleman2010JLTP}, the two phases are separated by an intermediate non-Fermi liquid phase (HS) with fermionic holon excitations (spin 0 and charge $+e$) and a partially enlarged electron Fermi surface. These may be compared to the experimental phase diagrams in CePdAl \cite{Zhao2019NatPhys} and Yb(Rh$_{1-y}$Ir$_y$)$_2$Si$_2$ \cite{Friedemann2009NatPhys}. The intermediate phase disappears under local approximations as shown in the inset of Fig.~\ref{fig4}a. Our method can be readily extended to other Kondo systems with geometric frustrations \cite{Wang2022SCPMAa}. For the triangular lattice, an effective gauge theory beyond the large-$N$ mean-field approximation can be derived for the intermediate phase and identify it as a fractionalized heavy fermion liquid with long-lived, heavy holon quasiparticles coupled to $\mathbb{Z}_2$ gauge fields \cite{WangJ2022PRB}. This provides a strong-coupling theory for the so-called metallic spin liquid in contrast to the weak-coupling FL$^*$ theory \cite{Senthil2003PRL,Senthil2004PRB,Sachdev2012JPCM}. In general cases, the holon state may be viewed as a parent state for other instabilities such as holon superconductivity or holon charge density wave. The former provides an additional pairing channel for heavy fermion superconductivity, and the latter breaks the translational symmetry and may be identified as a partial Kondo screening phase \cite{Motome2010PRL,Aulbach2015PRB}.

Unlike the AFM, the ferromagnetic (FM) quantum criticality was seldom reported in clean systems \cite{Steppke2013Science} and only recently observed in the stoichiometric compound CeRh$_6$Ge$_4$ \cite{Shen2020Nature}. Conventional theories predicted that the FM quantum phase transition in metallic systems should be first-order \cite{Belitz1999PRL,Belitz2012PRB}, which seems to be supported by experiments in many heavy fermion ferromagnets such as UGe$_2$ \cite{Aoki2019JPSJ}. To solve this issue, we have extended the Schwinger boson approach to study an anisotropic Kondo-Heisenberg model on the  square lattice \cite{Wang2022SCPMAb}. It was found that quasi-one-dimensionality may play a key role in causing the FM quantum criticality. Figure \ref{fig4}b compares the theoretical phase diagram with the experiments on CeRh$_6$Ge$_4$ and UGe$_2$ and we find good agreement with both compounds. The theoretical phase diagram is also supported by more accurate tensor network calculations \cite{Chen2022PRB}. Inside the FM, our calculations revealed a special spin-selective Kondo insulator (half metal) state featured with a magnetization plateau \cite{Li2010PRB,Peters2012PRL,Peters2012PRB,Golez2013PRB}.

The nonlocal Schwinger boson approach allows us to define a gauge-invariant quantity, the holon Fermi volume $V_{\rm FS}^\chi$, which varies between 0 and 1 and is associated with the size of the electron Fermi surface ($V_{\rm FS}^c$) through a generalized Luttinger sum rule: $NV_{\rm FS}^c=n_c+V_{\rm FS}^\chi$ \cite{Coleman2005PRB}. Here $N$ is the band degeneracy and $n_c$ is the electron density. It hence provides a measure of the $f$ electron itinerancy and may be regarded as a ``two-fluid order parameter". As in the two-fluid model, a notable prediction of this theory is that the electron Fermi surface can be partially enlarged at finite temperatures even on the magnetic side of the phase diagram, which differs from the prediction of a local QCP \cite{Si2010PSSB} but explains the ARPES observations in YbRh$_2$Si$_2$ \cite{Mo2012PRB,Kummer2015PRX} and CeRh$_6$Ge$_4$ \cite{Wu2021PRLa}. This suggests that how the Fermi surfaces evolve is closely associated with nonlocal spatial correlations. The nonlocal Schwinger boson approach indeed captures some essential physics of the Kondo lattice systems despite of its large-$N$ approximation.

Quantum phase transitions may also be driven by chemical substitution, which introduces chemical pressure, valence change, or disorder \cite{Miranda1996JPCM}. Replacing $f$-electron ions with their nonmagnetic counterparts can also change the Kondo lattice system into a dilute Kondo impurity system. The related physics is also very interesting but will not be discussed here \cite{Wei2017SR,Xie2017SR,Raczkowski2019PRL}.

\section{Quantum critical superconductivity}\label{secSC}
Unconventional superconductivity often emerges near the QCP. A salient feature of heavy fermion superconductivity is that it can appear at the border of many different competing orders, including magnetic or nonmagnetic ones \cite{Pfleiderer2009RMP,Mathur1998Nature,White2015PhyC,Yang2015AcPhySin,Li2021AcPhySin}. Multiband property is also a crucial ingredient. For over three decades, superconductivity in CeCu$_2$Si$_2$ was believed to be of $d_{x^2-y^2}$ wave based on the simplified one-band model \cite{Steglich1979PRL}, until two nodeless gaps were observed at very low temperatures in high quality samples in 2014 \cite{Kittaka2014PRL}. 

\begin{figure*}[t]
\centering\includegraphics[width=0.95\textwidth]{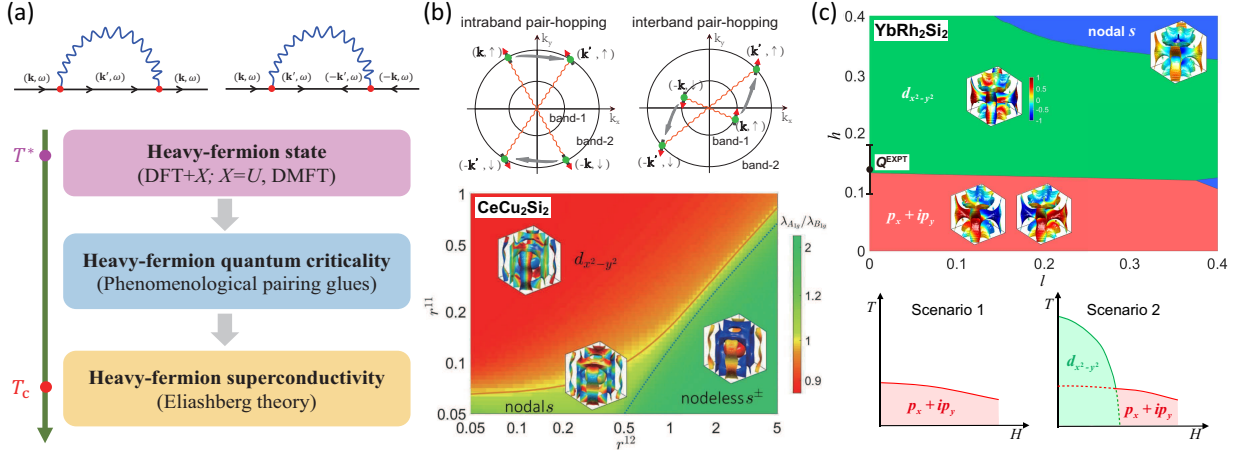}
\caption{A general phenomenological framework for heavy fermion superconductivity. (a) Illustration of the framework based on the Eliashberg theory with realistic band structures and quantum critical pairing glues \cite{Li2021AcPhySin}; (b) Theoretical phase diagram for CeCu$_2$Si$_2$ with both intraband and interband pair scatterings \cite{Li2018PRL}; (c) Superconducting phase diagram of YbRh$_2$Si$_2$ as a function of the propagation wave vector of the pairing glue, implying two candidate scenarios under the magnetic field \cite{Li2019PRB}. Figure adapted from \cite{Li2021AcPhySin,Li2018PRL,Li2019PRB}. Copyright 2018 and 2019 American Physical Society and 2021 Chinese Physical Society.}
\label{fig5}
\end{figure*}

To cover this richness, we have developed a general framework based on the Eliashberg theory \cite{Eliashberg1960SPJ}, with realistic band structures and quantum critical pairing glues as the input (see Fig.~\ref{fig5}a) \cite{Li2017CSB,Li2018PRL,Li2019PRB,Liu2019CPB,Sheng2022PRB}. Since the latter cannot be easily evaluated in theory, a phenomenological form of quantum critical fluctuations may be used \cite{Millis1990PRB},
\begin{equation}
V({\bf{q}},i\nu_n) = \frac{1}{1+[{\bf \xi}\cdot({\bf{q}}-{\bf{Q}})]^2+|\nu_n|/\omega_0},
\label{vertex}
\end{equation}
where $\nu_n$ is the bosonic Matsubara frequency, ${\bf\xi}$ is the anisotropic correlation length of quantum critical fluctuations, $\omega_0$ is the characteristic fluctuation energy, and ${\bf{Q}}$ is the propagation wave vector. This empirical form was first developed for cuprates and derived by expanding the spin interaction around ${\bf{Q}}$ \cite{Millis1990PRB,Monthoux1991PRL,Monthoux1992PRB,Monthoux1992PRL,Monthoux1993PRB,Monthoux1999PRB,Monthoux2001PRB,Monthoux2002PRB}. We have extended it to other pairing interactions in realistic materials \cite{Li2017CSB,Li2018PRL,Li2019PRB,Liu2019CPB,Sheng2022PRB}. The band structures may be either obtained from first-principles calculations \cite{Blaha2001Book,Kotliar2006RMP,Held2008JPCM,Haule2010PRB} or extracted from ARPES measurement. As discussed in section \ref{secQPT}, the Fermi surface should be nearly large in the normal state around the QCP. The predominant pairing state can then be solved using the Eliashberg equations. 

For CeCu$_2$Si$_2$, both band calculations and ARPES experiments have confirmed the presence of two Fermi surfaces (one electron-like and one hole-like) \cite{Zwicknagl1993PhyB,Ikeda2015PRL,Wu2021PRLb}. Studies using the above framework show that its superconductivity is $d_{x^2-y^2}$ wave if only the heavier electron Fermi surface is considered but could be $s^\pm$ wave when sufficient interband pair scattering between two Fermi surfaces is taken into account, as plotted in the theoretical phase diagram Fig.~\ref{fig5}b \cite{Li2018PRL}. This is similar to the pnictide superconductivity and provides a potential explanation of the observed two nodeless gaps \cite{Kittaka2014PRL,Kittaka2016PRB,Enayat2016PRB,Takenaka2017PRL,Yamashita2017SciAdv,Wang2019FP,Zhao2022PRB}. An alternative scenario is the so-called $d+d$ wave, which involves electron pairing between two hybridization bands and may thus require finite-momentum pairing due to the large direct hybridization gap \cite{Pang2018PNAS,Sichelschmidt2013JPCM}.

YbRh$_2$Si$_2$ is another example. This compound has been intensively studied for its type I AFM QCP \cite{Si2001Nature,Coleman2001JPCM,Gegenwart2008NatPhys,Stockert2011ARCMP,Abrahams2012PNAS,Steglich2014PM}. In 2016, superconductivity was detected below 2 mK inside the AFM phase \cite{Schuberth2016Science}, and stimulated immediate debate concerning its pairing mechanism. We find that the wave vector $\textbf{Q}$ of its magnetic fluctuations plays an essential role in determining the pairing symmetry \cite{Li2019PRB}. In experiment, $\textbf{Q}$ evolves from FM-like at high temperatures to an incommensurate AFM one at low temperatures \cite{Stock2012PRL}. As shown in Fig.~\ref{fig5}c, our  calculations indicate that the system is located on the border of $d_{x^2-y^2}$ singlet and $p_x+ip_y$ triplet pairing. Under magnetic field, the $d_{x^2-y^2}$ singlet solution may be suppressed and give way to the $p_x+ip_y$ triplet solution \cite{Li2019PRB}. The existence of two superconducting states (scenario 2) is confirmed very recently in experiment \cite{Nguyen2021NC}. 

The same framework may also be applied to other unconventional superconductors with quantum critical pairing glues. For example, in Sr$_2$RuO$_4$ \cite{Pustogow2019Nature}, our calculations suggest that a candidate $d_{x^2-y^2}+ig$ (pseudospin) singlet pairing state \cite{Kivelson2020npjQM} could naturally arise from the interplay of AFM, FM, and electric multipole fluctuations in the presence of the spin-orbit coupling \cite{Sheng2022PRB}. The success of our phenomenological framework indicates that combining quantum critical pairing glues and realistic electronic band structures can already explain the pairing mechanism in many unconventional superconductors without necessarily resorting to other more exotic assumptions.

\section{Material-specific properties}\label{secMat}
The generic physics discussed in previous sections or some distorted form should be present in all prototypical Kondo lattice systems, and the most notable examples have been given in each section. However, the richness of heavy fermion phenomena are closely associated with their rich variety of material-specific properties, which include but are not limited to the dimensionality, the valence, the orbital character, and so on. These properties not only affect how the generic properties behave, but may also cause other exotic physics beyond the standard Kondo lattice model. This section will introduce some of our recent material studies on these aspects.

\subsection{Dimensionality}
Dimensionality plays a crucial role in determining the ground state and quantum critical properties of heavy fermion materials. The quasi-one-dimensional Kondo lattice CeCo$_2$Ga$_8$ is located at a QCP under ambient pressure and zero magnetic field without tuning \cite{Wang2017npjQM}. It exhibits uniaxial hybridization at low temperatures \cite{Zheng2022PRB} and may therefore be viewed as a laboratory realization of the Kondo chain \cite{Cheng2019PRM}. As already discussed in section 4, the quasi-one-dimensional CeRh$_6$Ge$_4$ shows unusual FM quantum criticality and strange metallic behavior under pressure. In the recently-discovered triplet superconductor UTe$_2$ \cite{Ran2019Science}, our first-principles calculations revealed a two-leg ladder magnetic structure with frustrated inter-ladder interactions and quasi-two-dimensional hole and electron Fermi surfaces \cite{Xu2019PRL}, which have been confirmed by later experiments \cite{Duan2020PRL,Miao2020PRL,Aoki2022JPSJ} and could be important for understanding the superconducting pairing symmetry.

The layer compound NaYbSe$_2$ is a quantum spin liquid candidate \cite{Dai2021PRX} formed of Yb triangular lattices \cite{Jia2020CPL,Zhang2020arXiv}. It undergoes a structural transition at 11 GPa and an insulator-to-metal transition at 58.9 GPa. The high-pressure structure contains two inequivalent Yb layers of different Yb-Se distances. Furthering increasing pressure to above 103.4 GPa drives the system into superconductivity with the highest $T_c\approx 8\,$K \cite{Jia2020CPL}. Our density functional theory plus dynamical mean-field theory (DFT+DMFT) calculations confirmed that the insulator-to-metal transition is induced by gap close of conduction bands, so it becomes first a low-carrier Kondo system after the transition and then a heavy fermion metal above about 75 GPa \cite{Xu2022npjQM}. Its superconductivity emerges when the itinerant $f$ electron Fermi surfaces become well-nested on one of the inequivalent Yb layers, while the other Yb layer remains a spin liquid due to its large Yb-Se distance. NaYbSe$_2$ is probably the first heavy fermion superconductor discovered under  such high pressures and offers a rare case of coexistent spin liquid and superconductivity.

\subsection{Valence}
While we have mostly focused on the Kondo lattice systems with well-defined $f$ moments, valence change or fluctuations may also have important consequences \cite{Lawrence1981RPP,Riseborough2016RPP,Miyake2017PM}. One example is the first-order isostructural valence transition in YbInCu$_4$ \cite{Felner1986PRB,Mushnikov2015FNT}, which is similar to the famous  $\alpha$-$\gamma$ transition in Ce metal \cite{Johansson1974PM,Allen1982PRL}. Intensive investigations have been carried out but debates still exist concerning their underlying mechanism. Recently, we have carried out comparative studies of band structure calculations and ARPES measurements on both systems. In the Ce thin films, we observed dispersive $f$ electron quasiparticle bands around the Fermi energy when the interlayer distance is reduced \cite{Wu2021NC}, indicating a selective Mott delocalization in the thin films and implying that the $\alpha$-$\gamma$ transition in bulk Ce might be associated with a bandwidth-driven Mott transition of Ce 4$f$ orbitals rather than the Kondo volume collapse. In YbInCu$_4$ \cite{Felner1986PRB,Mushnikov2015FNT}, we found that the band hybridization was almost unchanged across the transition, despite that the $f$ electron bands shift downwards and the Yb$^{3+}$ hole bands touch the Fermi energy below the transition temperature. Comparison with DFT+DMFT calculations suggests a valence-driven selective Mott transition of 4$f$ electrons in YbInCu$_4$ \cite{Yang2022NSR}, contrary to the conventional Kondo scenario that expects an abrupt change of the hybridization strength  across the first-order transition. In CeRhGe$_3$, critical valence fluctuations were argued to be underlying its non-Fermi-liquid resistance and superconductivity under high pressure \cite{Wang2018PRB,Wang2019PRB}

Unusual properties may often appear in heavy fermion compounds with a local ionic configuration containing multiple $f$ electrons \cite{Cox1998AdvPhys}. Besides the well-known topological Kondo insulator as in the mixed-valence compound SmB$_6$ \cite{Dzero2010PRL,Lu2013PRL,Zhou2017SB}, the most notable example is the so-called ``hidden order" in URu$_2$Si$_2$ with a 5$f^2$ nominal configuration \cite{Palstra1985PRL,Mydosh2011RMP,Mydosh2014PM}. Many theories have been proposed but the mystery is still unresolved. Motivated by the studies of nematicity in pnictides, we have recently discovered an in-plane anisotropic response to the uniaxial pressure without necessarily breaking the rotational lattice symmetry in the hidden order state of URu$_2$Si$_2$ \cite{Wang2022CPL}. In the newly-synthesized single crystal NdFe$_2$Ga$_8$, a ``hidden order" transition was also observed, possibly associated with some yet-to-be-identified multipolar order \cite{WangC2021PRB,WangX2022PRB}. Electric quadrupole orders have been observed in Pr-based heavy fermion compounds such as PrOs$_4$Sb$_{12}$ with non-Kramers doublet for the crystal field ground state of Pr$^{3+}$, which may be responsible for their unusual superconductivity or other exotic properties \cite{Maple2007JMMM,Onimaru2016JPSJ}.

\subsection{Novel $d$- or $p$-electron heavy fermion systems}
Heavy fermion properties have also been reported in some $d$-electron correlated systems including LiV$_2$O$_4$ \cite{Kondo1997PRL}, CaCu$_3$Ir$_4$O$_{12}$, CaCu$_3$Ru$_4$O$_{12}$, and even infinite-layer nickelate superconductors, but their mechanism can be quite different. In CaCu$_3$Ir$_4$O$_{12}$ \cite{Cheng2013PRL}, we found that a mirror  symmetry could prohibit the nearest-neighbor hopping of Cu 3$d_{xy}$ electrons and make them more localized as in a Kondo lattice \cite{Liu2017arXiv}. But in CaCu$_3$Ru$_4$O$_{12}$ \cite{Kobayashi2004JPSJ}, the Cu 3$d_{xy}$ electrons are less correlated, so comparison between ARPES measurements and DFT+DMFT calculations revealed a gradual hybridization between emergent Cu 3$d_{xy}$ quasiparticle bands and the conduction bands with lowering temperatures \cite{Liu2020PRB}. 

Infinite-layer nickelate superconductors such as Nd$_{1-x}$Sr$_x$NiO$_2$ \cite{Li2019Nature} were proposed to be a low carrier density Kondo lattice system in the underdoped region, where the almost half-filled Ni 3$d_{x^2-y^2}$ electrons provide localized spins and a small fraction of them are transferred to the Nd 5$d$ bands as conduction electrons  \cite{Zhang2020PRB}. Incoherent Kondo scattering then dominates the transport properties and causes an anomalous upturn and then saturation in the resistivity \cite{Li2019Nature,Shao2022arXiv}. The proximity to a valence instability may be responsible for the charge order observed in recent experiments \cite{Rossi2022NatPhys,Kriger2022PRL,Tam2022NM}, which in turn promotes the self-doping and produces a condition for the Mott-Kondo scenario \cite{Chen2022arXiv}. The interplay of AFM fluctuations and Kondo hybridization could potentially support an exotic $(d+is)$-wave pairing state and yield multiple superconducting phases \cite{WangZ2020PRB}. The nickelate physics hence bridges those of heavy fermions and cuprates, which makes the nickelates a new class of unconventional superconductors with their own unique properties \cite{Yang2022FP,Gu2022Innovation}. 

Very recently, flat band systems such as van der Waals heterostructures \cite{Vano2021Nature} and magic-angle twisted bilayer graphene \cite{Song2022PRL} were also proposed to exhibit heavy fermion behaviors. All these greatly extend the scope of the heavy fermion researches.

\section{Conclusion}
To summarize, we have discussed four generic topics on heavy fermion physics from the development of phase coherence at high temperatures, to the two-fluid behavior below the coherence temperature, to the quantum phase transitions at zero temperature, to the rich superconductivity near the quantum critical point. For each topic, a different theoretical description has been constructed to capture its essential physics. In addition to these generic properties, we also discussed material-specific properties such as the dimensionality, the valence, and the orbital character, which also have important influence on experimental observations in real materials. In the pursuit of a better theory, it is important to distinguish the properties that are generic or material-specific. Although many questions are yet to be answered, in particular concerning the validity and applicability of the methods, a unified picture is seen to  emerge for heavy fermion physics following these recent progresses, and we are already facing a challenge to unify all of them into a single  consistent theory. 

\ack We thank all collaborators for stimulating discussions. This work was supported by the National Natural Science Foundation of China (NSFC Grants No. 12174429, No. 11974397, and No. 11774401), the National Key Research and Development Program of China (Grant No. 2022YFA1402200), and the Strategic Priority Research Program of the Chinese Academy of Sciences (Grant No. XDB33010100).

\end{document}